\documentclass{article}
\usepackage{amsmath}
\usepackage{amssymb}
\usepackage{theorem}

\newcommand{\Fac}{\operatorname{Fac}}
\newcommand{\Pal}{\operatorname{Pal}}

\title{Infinite words without palindrome}
\author{Jean Berstel, Luc Boasson, Olivier Carton, Isabelle Fagnot}

\begin{document}

\maketitle

\begin{abstract}
   We show that there exists an uniformly recurrent infinite word
   whose set of factors is closed under reversal and which has only
   finitely many palindromic factors.
\end{abstract}
 
\section{Notations}

For a finite word~$w = w_1\cdots w_n$, the \emph{reversal} of~$w$ is the
word $\tilde{w} = w_n\cdots w_1$.  This notation is extended to sets by
setting $F^\sim = \{\tilde{w} \mid w \in F\}$ for any set~$F$ of finite
words.  A word~$w$ is a \emph{palindrome} if $\tilde{w} = w$.  A set~$F$ of
finite words is closed under reversal if $F^\sim = F$.

For an infinite word~$x$, we denote respectively by $\Fac(x)$ and
$\Pal(x)$, the set of factors of~$x$ and the set of factors of~$x$ which
are palindrome.  

An infinite word is uniformly recurrent if each of its factors occurs
infinitely many times with bounded gap.  Equivalently, $x$ is uniformly
recurrent if for any integer~$m$, there is an integer~$n$ such that any
factor of~$x$ of length~$n$ contains all factors of~$x$ of length~$m$.

If $x$ uniformly recurrent and if $\Pal(x)$ is infinite, the set of factors
of~$x$ is closed under reversal, that is $\Fac(x) = \Fac(x)^\sim$.  In
the following examples, we show that the converse does not hold.

\section{Over a $4$-letter alphabet}

Let $A$ be the alphabet $A = \{0, 1, 2, 3\}$.  Define by induction the
sequence $(x_n)_{n\ge0}$ of words over~$A$ by $x_0 = 01$ and $x_{n+1} =
x_n23\tilde{x}_n$.  The first values are $x_1 = 012310$, $x_2 =
01231023013210$ and $x_3 = 012310230132102301231032013210$.  We denote
by~$x$ the limit of the sequence $(x_n)_{n\ge0}$.

We claim that the word~$x$ has the following properties
\begin{itemize} \itemsep0cm
\item $x$ is uniformly recurrent,
\item $\Fac(x)$ is closed under reversal : $\Fac(x)^\sim = \Fac(x)$,
\item $\Pal(x)$ is finite : $\Pal(x) = A$.
\end{itemize}
It can be easily shown by induction on~$n$ that there is a sequence
$(x'_i)_{i\ge1}$ of words from $\{23, 32\}$ such that 
\begin{displaymath}
  x_{p+n} = x_px'_1\tilde{x}_px'_2x_px'_3\tilde{x}_px'_4x_p\cdots 
            x_px'_{2^n-1}\tilde{x}_p.
\end{displaymath}
Since each factor of~$x$ is factor of~$x_n$ for $n$ large enough, the
word~$x$ is uniformly recurrent.  If $w$ is a factor of~$x_n$, then
$\tilde{w}$ is a factor of~$x_{n+1}$.  This shows that $\Fac(x)^\sim =
\Fac(x)$.  The word~$x$ belongs to $((01+10)(23+32))^\omega$.  Therefore,
it has no factor of the form $aa$ of $aba$ for $a,b \in A$.  This shows
that $\Pal(x) = A$.

\section{Over a $2$-letter alphabet}

Define the morphism $h$ from $A^*$ to~$\{0,1\}^*$ as follows
\begin{displaymath}
  h : 
  \left\{
  \begin{array}{l}
    0 \mapsto 101 \\
    1 \mapsto 1001 \\
    2 \mapsto 10001 \\
    3 \mapsto 100001
  \end{array}
  \right.
\end{displaymath}
Note the image of each letter is a palindrome and $h(\tilde{w}) = 
h(w)^\sim$. Let $y$ be the infinite word $h(x)$.  The beginning of~$y$ is
the following.  
\begin{displaymath}
  y = 101100110001100001100110110001100001101\cdots
\end{displaymath}
We claim that $y$ has the following properties
\begin{itemize} \itemsep0cm
\item $y$ is uniformly recurrent,
\item $\Fac(y)$ is closed under reversal : $\Fac(y)^\sim = \Fac(y)$,
\item $\Pal(y)$ is finite. 
\end{itemize}
Since $y$ is the image by a morphism of uniformly recurrent word, it is
also uniformly recurrent.  Since $h(\tilde{w})$ is the mirror image
of~$h(w)$ for each word~$w$, equality $\Fac(y)^\sim = \Fac(y)$ holds.
Each word~$w$ from $\Pal(y)$ is a factor of a word of the form $h(aub)$
where $u$ belongs to $\Pal(x)$ and $a,b \in A$.  Since $\Pal(x)$ is finite,
$\Pal(y)$ is also finite.

Define by induction the sequence $(z_n)_{n\ge0}$ of words over~$\{0,1\}$ by
$z_0 = 01$ and $z_{n+1} = z_n01\tilde{z}_n$.  The first values are $z_1 =
010110$, $z_2 = 01011001011010$ and $z_3 = 010110010110100101011010011010$.
We denote by~$z$ the limit of the sequence $(z_n)_{n\ge0}$.  Note that $z$
is also equal to $g(x)$ where the morphism~$g$ is given by $g(0) = g(2) =
0$ and $g(1) = g(3) = 1$.  We claim that the word~$z$ has the following
properties
\begin{itemize} \itemsep0cm
\item $z$ is uniformly recurrent,
\item $\Fac(z)$ is closed under reversal : $\Fac(z)^\sim = \Fac(z)$,
\item $\Pal(z)$ is finite.
\end{itemize}
The first two properties are proved as for~$x$.  We claim that each
word~$w$ in $\Pal(z)$ satisfies $|w| \le 12$.  Note that it suffices to
prove that $\Pal(z)$ contains no word of length $13$ or~$14$.  We prove by
induction on~$n$ that no palindrome of length $13$ or~$14$ occurs in~$z_n$.
An inspection proves that no palindrome of length $13$ or~$14$ occurs in
either $z_3 = z_201\tilde{z}_2$ or in $\tilde{z}_201z_2$.  For $n \ge 3$,
the word~$z_n$ can be factorized $z_n = z_2t_n\tilde{z}_2$ and the word
$z_{n+1}$ is equal to $z_2t_n\tilde{z}_201z_2\tilde{t}_n\tilde{z}_2$.
Since $z_2$ is of length~$14$ a palindrome of length $13$ or~$14$ which
occurs in~$z_{n+1}$ occurs either in~$z_n$ or in $\tilde{z}_201z_2$.
The result follows from the induction hypothesis.

\section{Links with paperfolding}

We point out a few links between the words we have introduced and
the so-called folding word. For a finite word~$w = w_1\cdots w_n$ over
$\{0,1\}$, denote by $\hat{w}$ the word $\bar{w}_n\cdots \bar{w}_1$ where
$\bar{0} = 1$ and $\bar{1} = 0$.  

Define by induction the sequence $(t_n)_{n\ge0}$ of words over~$\{0,1\}$ by
$t_0 = 0$ and $t_{n+1} = t_n0\hat{t}_n$.  We denote by~$t$ the limit of the
sequence $(t_n)_{n\ge0}$.  The set of factors of~$t$ is not closed under
reversal.  Indeed, the word $01000$ is a factor of~$t$ whereas $00010$ is
not.  The word~$y$ is equal to~$f(t)$ where the morphism~$f$ is given by
$f(0) = 01$ and $f(1) = 10$.

\end{document}